\documentclass[twocolumn,secnumarabic,amssymb,nobibnotes, aps, prapplied,10pt]{revtex4-2}

\usepackage{mathrsfs,amsmath,bm,graphicx}
\renewcommand{\arraystretch}{2}
\usepackage{verbatim}
\usepackage{upgreek}
\usepackage[dvipsnames]{xcolor}
\usepackage[normalem]{ulem}
\begin{document}


\title{Momentum-space geometric structure of helical evanescent waves and its implications on near-field directionality}


\author{Lei Wei}
\email{lei.wei@kcl.ac.uk}
\affiliation{Department of Physics, King's College London, Strand, London, WC2R 2LS, United Kingdom}
\author{Francisco J. Rodr\'{i}guez-Fortu\~{n}o}
\affiliation{Department of Physics, King's College London, Strand, London, WC2R 2LS, United Kingdom}

\date{\today}

\begin{abstract}
In this work, a momentum-space geometrical structure in helical evanescent electromagnetic waves is revealed. It is shown that for every helical evanescent wave on a helicity-dependent half tangent line in momentum space, the orientation of each of its field, spin, and Poynting vectors is the same. This geometric structure reveals itself as a remarkable relation between the far-field and near-field components of the angular spectrum. Any general evanescent wavevector is linked to two points on the $k_{\rho}=k_0$ circle of propagating wavevectors via two helicity-dependent tangent lines. Knowing the field on the $k_{\rho}=k_0$ circle of a general dipolar source is sufficient to determine its entire evanescent angular spectrum. Applying this concept, we gain insights into near-field directionality by showing that every zero in the angular spectrum is a helicity singularity where two half-tangent lines of opposite helicity intersect. A powerful method for synthetic design of near-field directional sources is also devised, using structured helical illumination to gain full control of the near-field directionality. The results provide a fundamental insight of helical evanescent waves and have implications in areas where chiral light-matter interaction plays a central role. 
\end{abstract}

\maketitle

\section{Introduction}
Chiral light-matter interaction \cite{AECohenPRL2010, IFCorbatonPRA2012, PLodahlNature2017} is becoming an increasingly important topic in recent years as evidenced on at least two fronts: its application in distinguishing chiral molecules from their mirror images (enantiomers) \cite{EHendryNNANO2010,AECohenSci2011,JSChoiPRA2012,IFCorbatonACSPhotonics2019} and its potential in building so-called chiral interfaces for quantum optics \cite{ARauschenbeutelSci2014, ARauschenbeutelPRX2015, KYBliokhSci2015, PLodahlNature2017}.

Enantiomers, a pair of organic or biological molecules with same chemical composition but opposite chirality, can have completely different chemical properties and biological functions. Being able to distinguish and separate them is critical in chemical, biological and pharmaceutical research and industry. Such structural information can be measured by circular dichroism (CD) spectroscopy, a method that uses chiral light to probe the molecular chirality based on the different extinction of the enantiomers in response to different handedness of the circularly polarised probing light. However, due to weak interactions between molecules and the probing light, the CD measurement is often challenging. To tackle this challenge, various concepts involving interference \cite{AECohenSci2011,MNVesperinasJOptics2017} and resonant nanostructures \cite{EHendryNNANO2010,IFCorbatonACSPhotonics2019} have been investigated in recent years to enhance the local optical field strength while maintaining chirality. This motivates the study on the design of chiral nanostructures \cite{EHendryNL2012,IFCorbatonACSPhotonics2015,IFCorbatonPRX2016} and structured light \cite{MNVesperinasPTRSA2017,KCKruiningOptica2018,SNechayevPRB2019} aimed at more effective chiral light-matter interaction in order to reach desired functionalities. More fundamentally, the conservation laws discovered in various versions since the 1960s regarding the helicity/chirality of electromagnetic waves \cite{DMLipkinJMP1964,CalkinMGAJP1965, IBBirulaJMP1981,KYBliokhPRA2011,SMBarnettPRA2012,TGPhilbinPRA2013} are given physical interpretations in the context of chiral light-matter interaction, and have been expanded to include sources \cite{MNVesperinasPRA2015, GNienhuisPRA2016}, nanostructures \cite{IFCorbatonPRL2013,LVPoulikakosACSPhotonics2016,SNechayevPRB2019N2} as well as general dispersive \cite{FAlpeggianiPRL2018} and lossy media \cite{JEVLozanoPRL2018}.

Helicity is a pseudoscalar resulting from the projection of the total angular momentum onto the linear momentum \cite{IFCorbatonPRA2012}, while the electromagnetic field spin is a pseudovector. A propagating circularly polarised plane wave has a well defined helicity of $\pm 1$ and a logintudinal spin, aligned with the Poynting vector and with the direction of phase propagation. In recent years, another type of spin---the transverse spin which arises in evanescent waves and in the fields of interfering plane waves \cite{KYBliokhPRA2012,PBanzerJEOS2013,FJRFortunoSci2013,KYBliokhPR2015,AAielloNPhotonics2015,SNechayevPRL2018} has drawn lots of attention. Different from the longitudinal counterpart, the transverse spin in a transverse-magnetic (TM or p-) or transverse electric (TE or s-) polarised evanescent wave has a helicity of 0 and exhibits a special spin-momentum locking property, where the orientation of the spin is transverse and locked relative to the direction of phase propagation \cite{ARauschenbeutelPRX2015, KYBliokhSci2015}. This property is widely considered as a manifestation of the spin-orbital interaction in optics \cite{FJRFortunoNPhoton2015,KYBliokhSci2015,ZJacobOptica2016}. Due to the general existence of transverse spin in evanescent waves at the interface of surfaces \cite{FJRFortunoSci2013, DOConnorNC2014}, waveguides \cite{ARauschenbeutelSci2014} and photonic crystals \cite{LKuipersSci2018}, access to the spin-momentum locking property unlocks new regimes in various fields like chiral quantum optics \cite{PLodahlNature2017}, topological photonics \cite{EWaksSci2018,LKuipersSci2018,CTChanArxiv2018, KYBliokhNC2019}, optical lateral forces \cite{FJRFortunoNC2015} as well as optical metrology \cite{MNeugebauerNL2019,ZXiPRL2018,AMartinezNL2017}. 

While optical helicity strongly relates to electric and magnetic duality \cite{IFCorbatonACSPhotonics2015,KYBliokh2PRL2014,IFCorbatonPRL2013,SMBarnettPRA2012,JOTrigoPRA2019}, most spin-momentum locking research \cite{FJRFortunoSci2013, DOConnorNC2014,ARauschenbeutelSci2014,LKuipersSci2018} investigate the coupling through electric spin only. That's understandable due to the scarcity of magnetic optical materials in nature. However in recent years, various works have shown that artificial optical magnetism can be induced by resonances in nanostructures \cite{AIKuznetsovSci2016}. For instance, high-index dielectric nanoparticles are known to support electric and magnetic modes and have been used to demonstrate induced electric and magnetic dipolar sources \cite{JMGeffrinNC2012,SPersonNL2013,YHFuNC2013}. The induced magnetic dipole enables the observation of magnetic transverse spin \cite{NeugebauerPRX2018} and the magnetic spin orbital interaction of light \cite{MWangLSA2018}. More interestingly, the construction of combined electric and magnetic dipolar sources \cite{RAlaeeOL2015,TCoenenNC2014,NFvanHulstNL2014,SNechayevPRA2019} shows ways to achieve near-field directionality that goes beyond spin-momentum locking \cite{MFPicardiPRL2018,MFPicardiLSA2019}. The generalized Huygens and Janus dipoles complement the spinning dipoles and can separately be related to the real and imaginary part of the complex Poynting vector and the spin of the local electromagnetic field. Besides, there also exists a type of helical dipoles \cite{IFCorbatonPRB2013,IFCorbatonACSPhotonics2015,MNVesperinasPRA2015,XZPuyaltoOL2013,XZPuyaltoNanoscale2016} where the electric dipole moment $\mathbf{p}$ and the magnetic dipole moment $\mathbf{m}$ fulfil $\mathbf{p}=\pm i\mathbf{m}/c_0$ with $c_0$ being the speed of light in vacuum. One should note that not all spinning dipoles are necessarily helical dipoles, neither a source needs to contain any spinning electric or magnetic dipole to be helical. As a special case, a helical source composed of collinear electric and magnetic dipoles has been demonstrated experimentally \cite{PBanzerOptica2018} where $\mathbf{p}\times\mathbf{m}^*$, $\mathbf{p}\times\mathbf{p}^*$ and $\mathbf{m}\times\mathbf{m}^*$ are all zero. In the context of helicity conservation, it is shown that for any helical dipolar source that fulfils $\mathbf{p}=\pm i\mathbf{m}/c_0$ including the collinear case, the radiation EM field of every angular spectrum component is circularly polarised and shares the same helicity as the source. This property brings new physical insights in understanding various optical phenomena, such as Kerker's condition for zero backscattering \cite{XZPuyaltoOL2013}. A dual dipolar Mie particle that fulfils Kerker's condition naturally induces a helical source when excited by a circularly polarised plane wave. To keep the same helicity as the incident light, circular polarization of the scattering field along the backward direction would need to be orthogonal to the incident one, and as a result, light coupling into the backscattering direction is not possible. \\

In this work, we study the helicity angular spectrum of dipolar sources including helical dipoles as well as sources that exhibit near-field directionality, \textit{e.g.} Janus dipoles, general Huygens dipoles and elliptical dipoles. Studying the near-field components of the angular spectrum, we will show that there exists an extraordinary geometrical structure in momentum space: helical evanescent waves with wavevectors lying on a helicity-dependent half tangent line share the same circularly polarised field vectors with the point of tangency on the $k_{\rho}=k_0$ circle. Thanks to this extraordinary geometrical structure of helical evanescent waves, the fields of the entire evanescent angular spectrum can be determined by merely knowing the fields on the $k_{\rho}=k_0$ circle. This knowledge enables an approach on synthetic design of near-field directional dipolar sources and is further applied to construct structured helical illumination to induce dipolar sources inside dual nanoparticles that enable angular tuning of near-field directionality. 

\section{Momentum-space geometric structures of helical evanescent waves}
We start with a detailed discussion on the characteristics of different polarization bases applied to the angular spectrum representation of a time harmonic electromagnetic field in free space, with an assumed $e^{-i\omega t}$ time dependence. The electric field $\bm{\mathcal{E}}(x,y,z)$ can be expanded in terms of spatial frequencies along an $xy$ plane perpendicular to any arbitrary $z$ direction, as:  
\begin{equation}\label{eq:angularspectrum}
\bm{\mathcal{E}}(x,y,z)=\iint \mathbf{E}(k_x,k_y) e^{i(k_xx+k_yy+k_zz)}\mathrm{d}k_x\mathrm{d}k_y.
\end{equation}
$\mathbf{E}(k_x,k_y)$ is the angular spectrum of field at the $z=0$ plane which represents the amplitude of a plane wave or evanescent wave with a wavevector $\mathbf{k}=(k_x,k_y,k_z)$ and $e^{ik_zz}$ is the propagator in reciprocal space \cite{LNovotnyBOOK}. Being a solution to the wave equation, the wave-vector fulfils $k_z=\pm\sqrt{k_0^2-k_{x}^2-k_{y}^2}$, where $k_0 =\omega/c_0$ is the wave number, $c_0$ is the speed of light in vacuum, and the $\pm$ sign refers to fields on the upper $(z>0)$ or lower $(z<0)$ half spaces, respectively, when the source is situated at $z=0$. The associated magnetic field angular spectrum can be determined via $\mathbf{H}(k_x,k_y)=\frac{1}{Z_0}\hat{\mathbf{k}}\times\mathbf{E}$ where $Z_0$ is the impedance of free space and $\hat{\mathbf{k}} = \mathbf{k}/k_0$.

The angular spectrum $\mathbf{E}(k_x,k_y)$ is a three dimensional vector, but the divergence-free condition imposed by Maxwell's equations on each angular component $\mathbf{k} \cdot \mathbf{E} = 0$ reduces the number of degrees of freedom to two, allowing it to be expanded into two transverse basis polarization vectors:
\begin{equation}\label{eq:psbasis}
\mathbf{E}(k_x,k_y)=E_p\hat{\mathbf{e}}_p+E_s\hat{\mathbf{e}}_s,
\end{equation}
\noindent where $\hat{\mathbf{e}}_p$ is the $p$-polarized (transverse magnetic or TM) and $\hat{\mathbf{e}}_s$ is the $s$-polarized (transverse electric or TE) basis vector. However, this basis is not unique. We may for instance expand the spectrum in the helical basis $\hat{\mathbf{e}}_{\pm}(k_x,k_y)=\frac{1}{\sqrt{2}}(\hat{\mathbf{e}}_p\pm i\hat{\mathbf{e}}_s)$:
\begin{equation}\label{eq:helicalbasis}
\mathbf{E}(k_x,k_y)=E_{+}\hat{\mathbf{e}}_{+}+E_{-}\hat{\mathbf{e}}_{-}.
\end{equation}
For simplicity reasons, we rewrite the wavevector from Cartesian coordinates $\mathbf{k}=(k_x,k_y,k_z)$ to cylindrical coordinates $\mathbf{k}=k_{\rho}\hat{\bm\uprho}+k_z\hat{\mathbf{z}}$. These two expressions are equivalent via the relations $\hat{\bm\uprho}=\frac{1}{k_{\rho}}(k_x,k_y,0)$, $\hat{\bm\upvarphi}=\frac{1}{k_{\rho}}(-k_y,k_x,0)$ and $k_\rho=\sqrt{k_x^2+k_y^2}$. The various polarization bases $\hat{\mathbf{e}}_p$, $\hat{\mathbf{e}}_s$ and $\hat{\mathbf{e}}_{\pm}$ can be expressed as: 
\begin{align}\label{pshelicalvector}
\hat{\mathbf{e}}_p(k_x,k_y)&=\frac{k_z}{k_0}\hat{\bm\uprho}-\frac{k_{\rho}}{k_0}\hat{\bm z}, \\\nonumber
\hat{\mathbf{e}}_s(k_x,k_y)&=\hat{\bm\upvarphi},\\\nonumber
\hat{\mathbf{e}}_{\pm}(k_x,k_y)&=\frac{1}{\sqrt{2}}\left(\frac{k_z}{k_0}\hat{\bm\uprho}\pm i\hat{\bm\upvarphi}-\frac{k_{\rho}}{k_0}\hat{\mathbf{z}}\right).
\end{align}

The advantages of helical basis have been demonstrated by various works \cite{KYBliokhPRA2010,IFCorbatonPRB2013,IFCorbatonACSPhotonics2015,MNVesperinasPRA2015,XZPuyaltoOL2013,XZPuyaltoNanoscale2016,MNVesperinasPTRSA2017} in studying chirality and far-field optical activities. Different from conventional Cartesian coordinate representation, the helicity basis is associated with a spin-1 monopole Berry curvature in momentum space \cite{KYBliokhPRA2010} which corresponds to a geometric phase of $2\pi$. Importantly, these basis vectors apply both to the propagating wave and to the evanescent wave region of the angular spectrum. The propagating region $(k_{\rho}\leq k_0)$ corresponds to real-valued wave-vectors, while the evanescent region $(k_{\rho}>k_0)$ corresponds to imaginary values of $k_z$. The circle $k_{\rho} = k_0$ corresponds to plane waves propagating within the $XY$ plane and it is the threshold between propagating and evanescent waves in the $(k_x,k_y)$ transverse momentum space. In this work we will focus on the evanescent region of the spectrum, and so, without loss of generality, we introduce $\gamma_z$ via the relation $k_z = i \gamma_z$, so that we can work with the real valued $\gamma_z = \pm \sqrt{k_{\rho}^2-k_0^2}$ for evanescent waves, with the $\pm$ sign accounting for fields in the $z>0$ or $z<0$ half-spaces as usual.

\begin{figure}[!htp]
\centering
\includegraphics[width=0.9\textwidth]{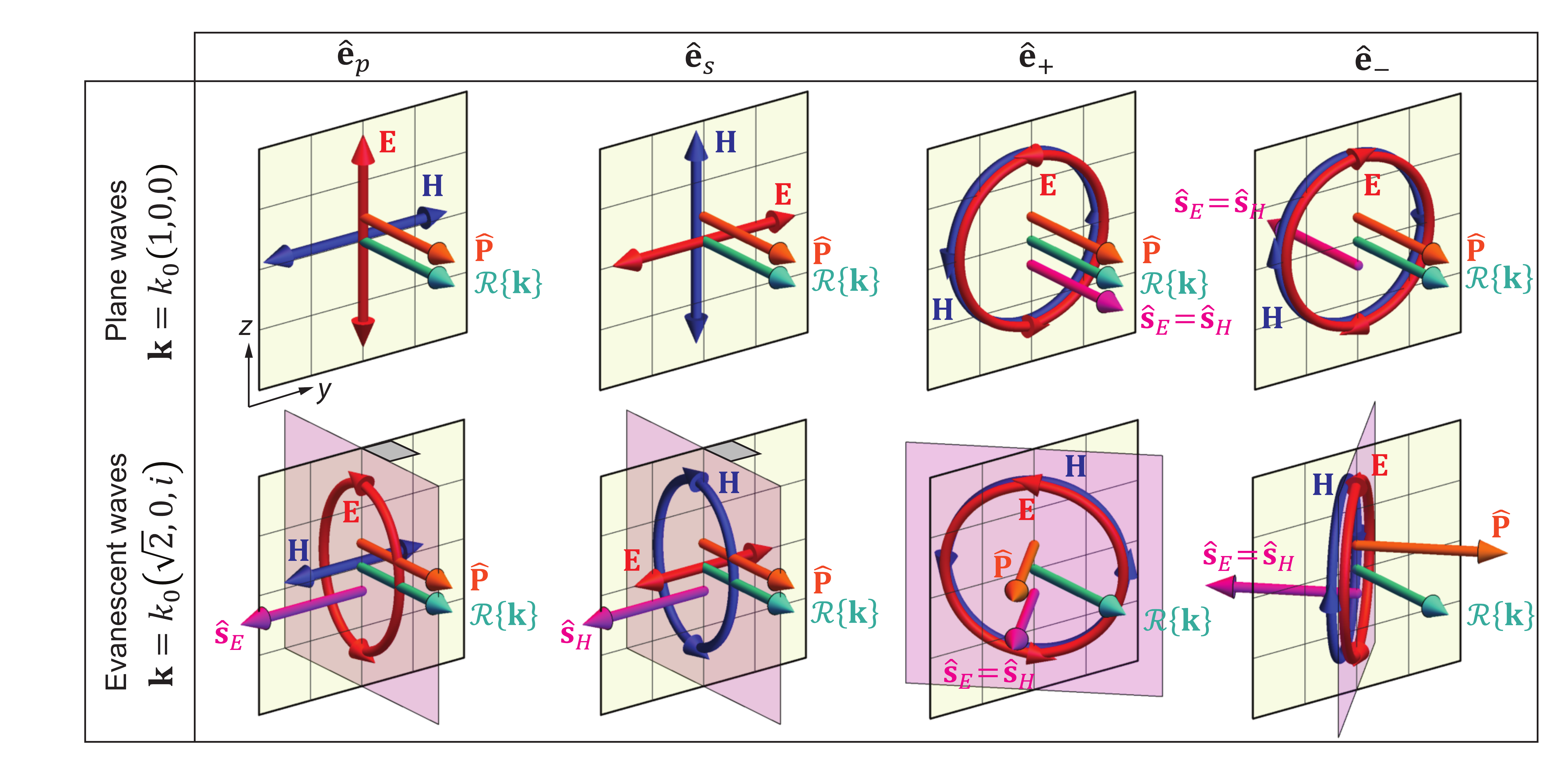}
\caption{The polarization ellipse, spin and Poynting vectors of propagating and evanescent electromagnetic waves with $\hat{\mathbf{e}}_p$, $\hat{\mathbf{e}}_s$ and $\hat{\mathbf{e}}_{\pm}$ polarised electric fields.}
\label{fig1l}
\end{figure}

In order to characterize the different polarization bases, we may extract a set of key physical properties from each basis vector: the helicity $\mathscr{H}$, the normalised Poynting vector $\hat{\mathbf{P}}$, and the normalised spin vectors of the electric field $\hat{\mathbf{s}}_E$ or the magnetic field $\hat{\mathbf{s}}_H$, defined as follows for each angular spectrum component \cite{KYBliokhPR2015}:
\begin{align}\label{eq:spinPoyntingHelicity}
&\hat{\mathbf{s}}_E(k_x,k_y)=\frac{\Im\left\{\mathbf{E}^*\times\mathbf{E}\right\}}{|\mathbf{E}|^2},\,\,\hat{\mathbf{s}}_H(k_x,k_y)=\frac{\Im\left\{\mathbf{H}^*\times\mathbf{H}\right\}}{|\mathbf{H}|^2},\\\nonumber
&\hat{\mathbf{P}}(k_x,k_y)=\frac{\Re\left\{\mathbf{E}^*\times\mathbf{H}\right\}}{|\mathbf{E}||\mathbf{H}|},\,\,\mathscr{H}(k_x,k_y)=\frac{2\Im\left\{\mathbf{E}\cdot Z_0\mathbf{H}^{*}\right\}}{|\mathbf{E}|^2+|Z_0\mathbf{H}|^2}.
\end{align}
The calculation of these physical properties for each of the basis vectors is shown in Table \ref{table1}. To visually understand the results, Fig. \ref{fig1l} illustrates the polarization ellipse and the calculated physical properties $\hat{\mathbf{P}}$ and $\hat{\mathbf{s}}$ for each basis vector polarization. The top row corresponds to propagating plane waves: the $\hat{\mathbf{e}}_p$ and $\hat{\mathbf{e}}_s$ polarizations correspond to linearly polarized waves, with their Poynting vector parallel to their wave-vector indicating the direction of energy flow, and no spin vectors. Their helicity is $\mathscr{H} = 0$. The $\hat{\mathbf{e}}_+$ and $\hat{\mathbf{e}}_-$ plane wave polarizations correspond to the two circularly polarized plane waves, also with their Poynting vector parallel to their wave-vector, but with a non-zero spin parallel or anti-parallel to them. Their helicity is $\mathscr{H} = \pm 1$ and the spin and Poynting vectors follow the relation $\hat{\mathbf{s}}=\mathscr{H} \hat{\mathbf{P}}$.

\begin{table}[!htbp]
\centering
\caption{Helicity $\mathscr{H}$, normalised Poynting vector $\hat{\mathbf{P}}$ and spin vectors $\hat{\mathbf{s}}_E$ of the electric field and $\hat{\mathbf{s}}_H$ magnetic field of evanescent waves for $\hat{\mathbf{e}}_p$, $\hat{\mathbf{e}}_s$ and $\hat{\mathbf{e}}_{\pm}$ polarised fields. See TABLE \ref{table2} for the corresponding properties in propagating waves.}
\begin{tabular}{|c|c|c|c|}
\hline
&$\hat{\mathbf{e}}_p$ (TM)&$\hat{\mathbf{e}}_s$ (TE) &$\hat{\mathbf{e}}_{\pm}$ (Helical) \\
\hline
$\hat{\mathbf{s}}_E$&$-\frac{2k_{\rho}\gamma_z}{k_{\rho}^2+\gamma_z^2}\hat{\bm{\upvarphi}}$&0&$\pm\frac{k_0}{k_{\rho}}\hat{\bm{\uprho}}-\frac{\gamma_z}{k_{\rho}}\hat{\bm{\upvarphi}}$\\
\hline
$\hat{\mathbf{s}}_H$&0&$-\frac{2k_{\rho}\gamma_z}{k_{\rho}^2+\gamma_z^2}\hat{\bm{\upvarphi}}$&$\pm\frac{k_0}{k_{\rho}}\hat{\bm{\uprho}}-\frac{\gamma_z}{k_{\rho}}\hat{\bm{\upvarphi}}$\\
\hline
$\hat{\mathbf{P}}$&$\frac{k_{\rho}}{\sqrt{k^2_{\rho}+\gamma_z^2}}\hat{\bm{\uprho}}$&$\frac{k_{\rho}}{\sqrt{k^2_{\rho}+\gamma_z^2}}\hat{\bm{\uprho}}$&$\frac{k_0}{k_{\rho}}\hat{\bm{\uprho}}\mp\frac{\gamma_z}{k_{\rho}}\hat{\bm{\upvarphi}}$\\
\hline
$\mathscr{H}$&0&0&$\pm1$\\
\hline
\end{tabular}
\label{table1}
\end{table}

\onecolumngrid
\widetext
\twocolumngrid
The bottom row of Fig. \ref{fig1l} illustrates the polarization of evanescent waves. As an example we choose a wave-vector which is propagating in the $x$ direction with $k_x = \sqrt{2} k_0$, and a corresponding imaginary $k_z = ik_0$ determining the direction of exponential decay. As is well known, the $\hat{\mathbf{e}}_p$ and $\hat{\mathbf{e}}_s$ evanescent polarizations acquire a transverse electric or magnetic spin, respectively, perpendicular to the wave-vector, and independent of the polarization \cite{KYBliokhPR2015,ZJacobOptica2016}. However, their helicity (which is related to the projection of the spin in the direction of power flow) is still $\mathscr{H} = 0$, because the Poynting vector is parallel to the real part of their wave-vector. Perhaps less well known is the behaviour of the evanescent $\hat{\mathbf{e}}_+$ and $\hat{\mathbf{e}}_-$ polarizations: their helicity is still given by $\mathscr{H} = \pm1$, and hence can be properly referred to as helical evanescent waves, and the polarization of their electric and magnetic fields are always perfectly circular, as for plane waves. Just like circularly polarised plane waves in free space, the normalised spin and Poynting vectors of a helical evanescent wave follow the relation $\hat{\mathbf{s}}=\mathscr{H} \hat{\mathbf{P}}$, meaning that they are still parallel or anti-parallel to one another. However, remarkably, both vectors are at an angle to the real part of the wave-vector. The tilt of the spin vectors from the direction of phase propagation is a result of the polarization-independent transverse spin of evanescent waves, while the Poynting vector of a helical evanescent wave acquires an helicity-dependent transverse component, as shown in Table \ref{table1}. This is clearly seen in Fig. \ref{fig1l} because the plane of circular polarization is not perpendicular to the propagation direction; instead, the polarization plane tilts, and it does so in different directions for the two opposite helicities. The existence of this rather 'unusual' transverse Poynting vector has actually been predicted alongside the discovery of Fedorov-Imbert transverse shift of elliptically polarised beams under total internal reflection \cite{CImbert1972,FedorovJO2013} and still invites discussions on its actual role in the transverse beam shift even to this day \cite{KYBliokhNC2014,KYBliokhNPhys2016}. In a recent work \cite{KYBliokhNC2014}, this transverse component of Poynting vector predicted by Fedorov is shown to be a `virtual' Belinfante's momentum.
 
\begin{figure}[!htp]
\centering
\includegraphics[width=0.45\textwidth]{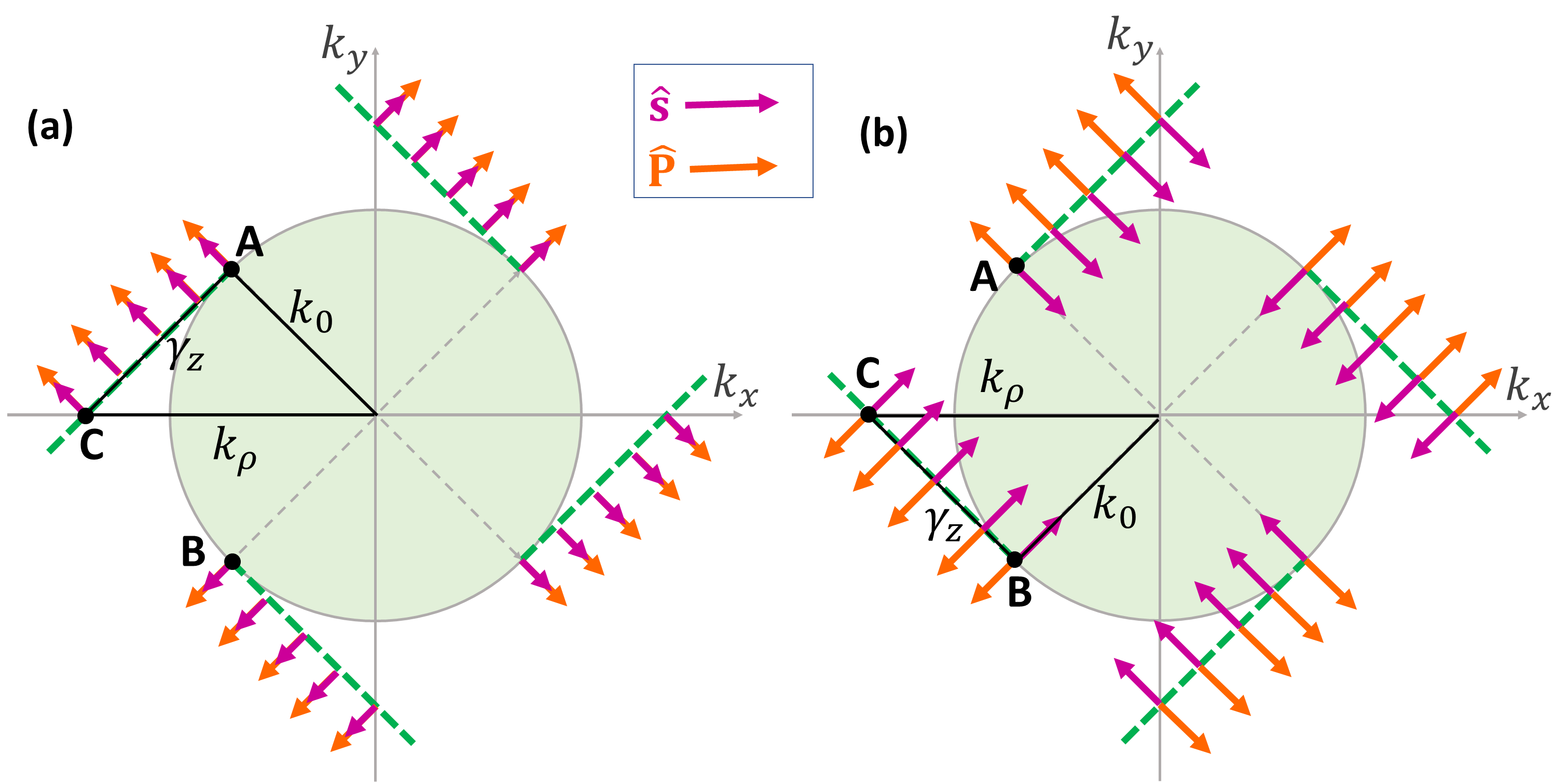}
\caption{Momentum-space (upper half space) geometrical structures of the normalised spin $\hat{\mathbf{s}}$ and Poynting vector $\hat{\mathbf{P}}$ of helical evanescent waves with field polarization vectors (a) $\hat{\mathbf{e}}_{+}(k_x,k_y)=\frac{1}{\sqrt{2}}\left[i\left(\frac{\gamma_z}{k_0}\hat{\bm\uprho}+\hat{\bm\upvarphi}\right)-\frac{k_{\rho}}{k_0}\hat{\mathbf{z}}\right]$ and (b) $\hat{\mathbf{e}}_{-}(k_x,k_y)=\frac{1}{\sqrt{2}}\left[i\left(\frac{\gamma_z}{k_0}\hat{\bm\uprho}-\hat{\bm\upvarphi}\right)-\frac{k_{\rho}}{k_0}\hat{\mathbf{z}}\right]$.}
\label{fig2}
\end{figure}

While the properties described above are known, in this work we specifically address the tilt angle of the spin and Poynting vector with respect to the real wavevector direction in helical evanescent waves. As the transverse wavevector $k_{\rho}$ increases from the free-propagating wave $k_{\rho}=k_0$ to a strongly evanescent wave with large $k_{\rho}$, both the spin and Poynting vector associated to $\hat{\mathbf{e}}_\pm$ are gradually rotating from being longitudinal $(\hat{\bm\uprho})$ for plane waves, to being transverse $(\pm\hat{\bm\upvarphi})$ in the limit of $k_{\rho} \to \infty$. One may then ask what is the exact tilt angle for each value of $k_{\rho}$. The answer is evidently contained in the mathematical expressions of the vectors given in Table \ref{table1}, but a visual geometrical interpretation reveals itself when we plot the two vectors $\hat{\mathbf{P}}$ and $\hat{\mathbf{s}}$ for various positions along the evanescent region of the spectrum, for the two helical polarisations, as shown in Fig. \ref{fig2}. The key feature, being the main novelty of this work, is the following: any $\hat{\mathbf{e}}_{\pm}$ polarised evanescent waves whose wavevector lies on the half tangent line starting from a point on the $k_{\rho}=k_0$ circle and directed towards the $\pm\hat{\bm \upvarphi}$ direction, respectively, share the same tilt in their circular polarization, and hence the same normalised spin and Poynting vectors. The above result can also be proven mathematically from Eqs. (\ref{pshelicalvector}), with the substitution $k_z = i \gamma_z$, together with Table \ref{table1} and basic trigonometry applied to the right angled triangles highlighted in Fig. \ref{fig2}. Globally over the momentum space, these helicity-dependent half tangent lines give rise to a chiral structure of aligned polarization, spin and Poynting vectors in the evanescent part of the angular spectrum. The circularly polarised basis vectors of the two opposite helicities lead to a structure with opposite chirality in the momentum space. Also, the angular spectra of the fields at the two different half-spaces ($z>0$ or $z<0$) have opposite chirality for the same helicity, owing to the change in sign of $\gamma_z$.

This geometric interpretation has very interesting consequences for near-field directionality, as discussed in the second half of this manuscript. Most importantly, a remarkable relation between the far-field and near-field components of the angular spectrum is revealed. As shown in Fig. \ref{fig2}, any general evanescent wavevector can be linked to two $k$-points on the $k_{\rho}=k_0$ circle of propagating wavevectors via two half tangent lines. For example, take the $k$-point labelled as \textbf{C}: the basis vector $\hat{\mathbf{e}}_{+}(\mathbf{C})$ at this point is the same as the basis vector $\hat{\mathbf{e}}_{+}(\mathbf{A})$ at the $k$-point \textbf{A} apart from a real-valued scaling factor $\hat{\mathbf{e}}_{+}(\mathbf{C})=\frac{k_{\rho}}{k_0}\hat{\mathbf{e}}_{+}(\mathbf{A})$, while \textbf{C} shares the same basis vector $\hat{\mathbf{e}}_{-}$ with $k$-point \textbf{B} apart from the same scaling factor $\hat{\mathbf{e}}_{-}(\mathbf{C})=\frac{k_{\rho}}{k_0}\hat{\mathbf{e}}_{-}(\mathbf{B})$. The fact that the evanescent helical basis $\hat{\mathbf{e}}_{\pm}$ has a constant polarization basis vector orientation throughout the corresponding half tangent line has a very important consequence on the angular spectrum of dipolar sources. Consider a time harmonic dipolar source positioned at the origin with an electric dipole moment $\mathbf{p}$ and a magnetic dipole moment $\mathbf{m}$. The angular spectrum of its radiation field projected to TM/TE basis is well known \cite{LNovotnyBOOK} and may be compactly expressed as \cite{MFPicardiPRB2017,MFPicardiPRL2018,JEVLozanoPRApp2019}:
\begin{align} \label{eq:dipoleangularspectrumsp}
E_p(k_x,k_y)&\propto\left(\mathbf{p}-\hat{\mathbf{k}}\times\frac{\mathbf{m}}{c_0}\right)\cdot\hat{\mathbf{e}}_p,\\ \nonumber
E_s(k_x,k_y)&\propto\left(\mathbf{p}-\hat{\mathbf{k}}\times\frac{\mathbf{m}}{c_0}\right)\cdot\hat{\mathbf{e}}_s.
\end{align}
We may change this expansion into the helical basis, via $E_\pm = \frac{1}{\sqrt{2}}(E_p \mp i E_s)$, arriving at:
\begin{align}\label{eq:dipoleangularspectrumhelical}
E_{\pm}(k_x,k_y)&\propto\left(\mathbf{p}\pm i\frac{\mathbf{m}}{c_0}\right)\cdot\hat{\mathbf{e}}_{\mp}
\end{align}
Notice that the term $\left(\mathbf{p}-\hat{\mathbf{k}}\times\frac{\mathbf{m}}{c_0}\right)$ which appears in the $\left\lbrace \hat{\mathbf{e}}_{s}, \hat{\mathbf{e}}_{p} \right\rbrace$ basis is a function of $\hat{\mathbf{k}}$ while the term $\left(\mathbf{p}\pm i\frac{\mathbf{m}}{c_0}\right)$ appearing in the less usual helical $\left\lbrace \hat{\mathbf{e}}_{+}, \hat{\mathbf{e}}_{-} \right\rbrace$ basis is independent of $\hat{\mathbf{k}}$, and hence is constant throughout the spectrum. This implies that the only variation of the amplitudes $E_+(k_x,k_y)$ and $E_-(k_x,k_y)$ throughout the full angular spectrum is provided by the basis vectors $\hat{\mathbf{e}}_{\mp}$. Therefore, any existing geometric structures in the angular spectrum of the evanescent helical basis vectors $\hat{\mathbf{e}}_{\mp}$, such as the tangent lines described above, can be directly applied to the field angular spectra $E_{\pm}$ too. In the same way that the helical basis vectors $\hat{\mathbf{e}}_{\pm}$ in the evanescent part of the angular spectrum can be linked to two points on the $k_{\rho}=k_0$ circle via half tangent lines, the field amplitude $E_{\pm}$ of any evanescent angular spectrum of a dipolar source can be uniquely determined by the spectrum amplitude at two points on the $k_{\rho}=k_0$ circle. Take, for example, the evanescent wave with the wavevector at $k$-point \textbf{C}: its $E_{-}$ coefficient can be determined by the $E_{-}$ coefficient at $k$-point \textbf{A} while its $E_{+}$ coefficient can be determined by the $E_{+}$ coefficient at $k$-point \textbf{B}. In other words, knowing the fields on the entire $k_{\rho}=k_0$ circle of any general dipolar source, the fields of its entire evanescent angular spectrum can be determined. This is the result of the general and universal tangent-line feature of helical evanescent waves and can be detected in Fourier polarimetry measurements \cite{AFKoenderinkLSA2018,PBanzerOptica2018,NeugebauerSciAdv2019}. In particular, the tangent line feature described in this work can be observed, without mention but present in the figures, in recent experiments \cite{PBanzerOptica2018,NeugebauerSciAdv2019}.

\section{Near-field directional dipolar sources}
This momentum space geometrical structure of helical evanescent waves has many great implications and forms the basis for the study of near-field directional dipolar sources and the construction of such sources using structured helical illumination discussed below. For instance, if either coefficient $E_{\pm}$ of the helical field components for a wavevector on the $k_{\rho}=k_0$ circle is zero, the same null coefficient will apply to all field components with the same helicity whose wavevector lies along the corresponding half tangent line. If $E_{+}=0$ on the $k_{\rho}=k_0$ circle then it follows from Eq. (\ref{eq:dipoleangularspectrumhelical}) that $E_{+}=0$ is zero on every wavevector on the half tangent line starting from the tangent point along $-\hat{\bm \upvarphi}$ in the momentum space for the fields in the upper half-space ($z>0$). Alternatively, $E_{-}=0$ on the $k_{\rho}=k_0$ circle will apply to any wavevector on the half tangent line starting from the tangent point along $+\hat{\bm \upvarphi}$ in the momentum space for the fields in the upper half-space ($z>0$).

\begin{figure}[!htp]
\centering
\includegraphics[width=0.45\textwidth]{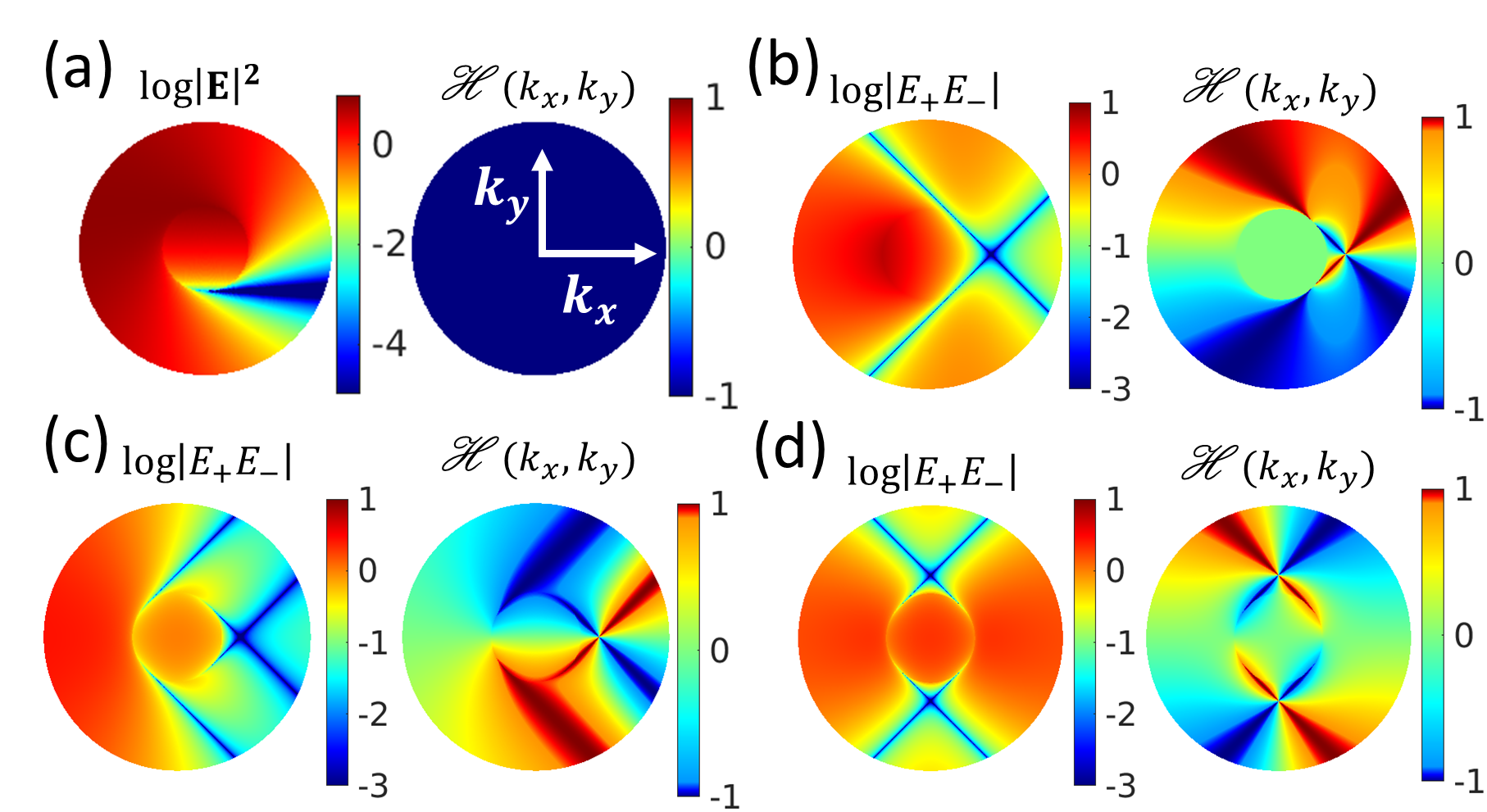}
\caption{(a) Angular spectrum (in the upper half space $z\geq0$) of energy and helicity for the source composed of an electric dipole $\mathbf{p}=\hat{\mathbf{x}}+i\hat{\mathbf{z}}$ and a magnetic dipole $\mathbf{m}=c_0(i\hat{\mathbf{x}}-\hat{\mathbf{z}})$; Angular spectrum (in the upper half space $z\geq0$) of the field coefficients $\log|E_{+}E_{-}|$ and helicity $\mathscr{H}$ of the sources:(b) $\mathbf{p}=\hat{\mathbf{z}}$ and $\mathbf{m}=c_0(\sqrt{2}\hat{\mathbf{y}})$; (c) $\mathbf{p}=\hat{\mathbf{x}}+i/\sqrt{2}\hat{\mathbf{z}}$ and $\mathbf{m}=0$; (d) $\mathbf{p}=\hat{\mathbf{x}}$ and $\mathbf{m}=c_0(i\hat{\mathbf{y}})$;}
\label{fig2:HAM_NFD}
\end{figure}

As an example, consider the source composed of a magnetic dipole $\mathbf{m}=c_0(i\hat{\mathbf{x}}-\hat{\mathbf{z}})$ and an electric dipole $\mathbf{p}=\hat{\mathbf{x}}+i\hat{\mathbf{z}}$. Such source fulfils the relation $\mathbf{p}=-i\mathbf{m}/c_0$. It is evident from Eq. (\ref{eq:dipoleangularspectrumhelical}) that its field is purely $\hat{\mathbf{e}}_{-}$ polarised in the entire momentum space, with its amplitude determined by $E_{-}\propto\left(\mathbf{p}- i\frac{\mathbf{m}}{c_0}\right)\cdot\hat{\mathbf{e}}_{+}$. On the $k_{\rho}=k_0$ circle, this corresponds to $\mathrm{E_{-}}\propto(1-\hat{\mathbf{x}}\cdot\hat{\bm \upvarphi})$. From the discussion above, $E_{-}=0$ at $\varphi=270^{\circ}$ on the $k_{\rho}=k_0$ circle, and this zero will extend to any wavevector on the half tangent line along $\hat{\bm \upvarphi}$ starting from this $k$ point. As the radiation field is solely determined by $E_{-}$, the total radiation into these wavevectors are completely suppressed as shown in Fig. \ref{fig2:HAM_NFD}(a).

The same analysis can be applied to characterize any general dipolar source that exhibits near-field directionality. Consider the following three types of known directional dipolar sources: Fig. \ref{fig2:HAM_NFD}(b) corresponds to a generalized Huygens dipole with an electric dipole $\mathbf{p}=\hat{\mathbf{z}}$ and a magnetic dipole $\mathbf{m}=c_0(\sqrt{2}\hat{\mathbf{y}})$; Fig. \ref{fig2:HAM_NFD}(c)  corresponds to an elliptical electric dipole $\mathbf{p}=\hat{\mathbf{x}}+i/\sqrt{2}\hat{\mathbf{z}}$ and $\mathbf{m}=0$; and Fig. \ref{fig2:HAM_NFD}(d) corresponds to a Janus dipole with an electric dipole $\mathbf{p}=\hat{\mathbf{x}}$ and a magnetic dipole $\mathbf{m}=c_0(i\hat{\mathbf{y}})$. For the generalized Huygens dipole shown in Fig. \ref{fig2:HAM_NFD}(b), at $k$ points with $\varphi=45^{\circ}$ and $\varphi=315^{\circ}$ on the $k_{\rho}=k_0$ circle, $E_{+}$ and $E_{-}$ are zero. Starting from each point, the evanescent waves on the half tangent line along $\hat{\bm \upvarphi}$ are helical with $\mathscr{H}=+1$ as $E_{-}=0$, while the evanescent waves on the half tangent line along $-\hat{\bm\upvarphi}$ are helical with $\mathscr{H}=-1$ as $E_{+}=0$. At the point of $k_{\rho}=\sqrt{2}k_0$ and $\varphi=0^{\circ}$, the two half tangent lines with opposite helicity intersect. As a result, at this point both $E_{+}$ and $E_{-}$ are zero, leading to zero total field at this wavevector, corresponding to the known Huygens dipole near field directionality. For the elliptical electric dipole shown in Fig. \ref{fig2:HAM_NFD}(c), there are four wavevectors on the $k_{\rho}=k_0$ circle whose fields are circularly polarized (and therefore have either $E_{+}$ or $E_{+}$ equal to zero): two at $\varphi=45^{\circ}$ and $\varphi=135^{\circ}$ with $\mathscr{H}=-1$ and the other two at $\varphi=225^{\circ}$ and $\varphi=315^{\circ}$ with $\mathscr{H}=+1$. Two of the corresponding half tangent lines of opposite helicity intersect at the point of $k_{\rho}=\sqrt{2}k_0$ and $\varphi=0^{\circ}$, leading to a zero total field at this wavevector, again corresponding to the known near-field directionality of elliptical dipoles. The Janus dipole in Fig. \ref{fig2:HAM_NFD}(d) also exhibits four wavevectors on the $k_{\rho}=k_0$ circle whose fields are circularly polarized, but in a different angular ordering: two at $\varphi=45^{\circ}$ and $\varphi=225^{\circ}$ with $\mathscr{H}=+1$ and the other two at $\varphi=135^{\circ}$ and $\varphi=315^{\circ}$ with $\mathscr{H}=-1$. The corresponding half tangent lines of opposite helicity lead to two crossing points and thus zero total field: one at $k_{\rho}=\sqrt{2}k_0$ and $\varphi=90^{\circ}$ and the other at $k_{\rho}=\sqrt{2}k_0$ and $\varphi=270^{\circ}$. This corresponds to the known off-coupling state of the Janus dipole.

The directional dipolar sources in Fig. \ref{fig2:HAM_NFD}(b)-(d) show that the near-field zero radiation wavevectors explored in prior works are also momentum space helicity singularities at the intersection of tangent lines. Knowing the wavevectors on the $k_{\rho}=k_0$ circle whose coefficient $E_{+}$ or $E_{-}$ is zero, the wavevectors of null radiation in the near-field angular spectrum can be determined thanks to the unique tangent line property of helical evanescent waves. 

\section{Constructing new dipoles and Structured helical illumination}
This knowledge is not only useful to characterize known near-field directional dipolar sources, but can also be used to devise a powerful method to find new dipolar sources for designated null-radiation wavevectors and further design structured helical illumination to excite such sources. 

\begin{figure}[!htp]
\centering
\includegraphics[width=0.45\textwidth]{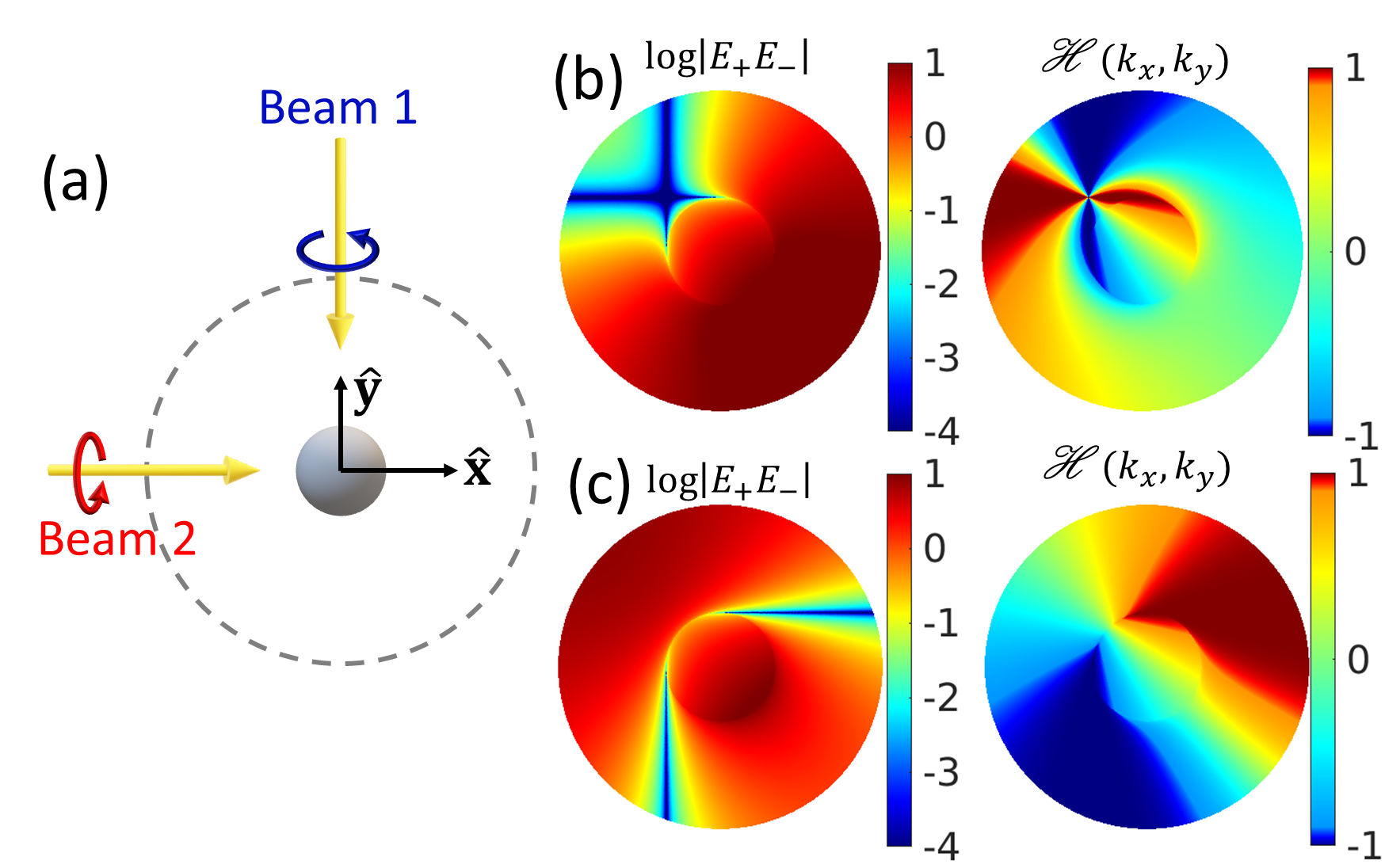}
\caption{(a). A dual dipolar nanoparticle illuminated by two helical beams; The scattering field and helicity angular spectrum (b) in the upper half space and (c) in the lower half space by induced dipole $\mathbf{p}_1+\mathbf{p}_2$ and $\mathbf{m}_1+\mathbf{m}_2$, where $\mathbf{p}_1=-i\hat{\mathbf{x}}-\hat{\mathbf{z}}$ and $\mathbf{m}_1=ic_0 \mathbf{p}_1$ is the induced helical dipole by beam 1 alone, $\mathbf{p}_2=i\hat{\mathbf{y}}-\hat{\mathbf{z}}$ and $\mathbf{m}_2=-ic_0 \mathbf{p}_2$ is the induced helical dipole by beam 2 alone.}
\label{fig3:HS1}
\end{figure}

Consider the case in Fig. \ref{fig3:HS1} where we want to find a dipolar source whose radiation into the wavevector $k_{\rho}=\sqrt{2}k_0$ at $\varphi=135^{\circ}$ in the upper half space to be zero. This desired wavevector is the crossing point of two half tangent lines, one starting from the $k$ points on the $k_{\rho}=k_0$ circle at $\varphi=90^{\circ}$ along $\hat{\bm\upvarphi}$ and the other half tangent line starting from the $k$ point on the $k_{\rho}=k_0$ circle at $\varphi=180^{\circ}$ along $-\hat{\bm\upvarphi}$. Based on the properties of helical evanescent waves, we know that the desired dipole source must have $E_{-}=0$ on the $k$ point at $\varphi=90^{\circ}$ of the $k_{\rho}=k_0$ circle and $E_{+}=0$ on the $k$ point at $\varphi=180^{\circ}$ of the $k_{\rho}=k_0$ circle. Both conditions are necessary and sufficient for the field amplitude at the designated wavevector to be zero. For the proposed method to work, one will need to involve two key elements as exemplified by the system in Fig. \ref{fig3:HS1}(a): one is helical beam illumination and the other is a dual dipolar nanoparticle. A nanoparticle with electric and magnetic dipole polarizabilities $\alpha_{\mathrm{e}}=\frac{i6\pi\epsilon_0}{k^3_0}a_1$ and $\alpha_{\mathrm{m}}=\frac{i6\pi}{k^3_0}b_1$ is dual when the ED and MD Mie coefficients are equal $a_1=b_1$. Such nanoparticles have some unique properties to serve our purpose. First of all, dual dipolar nanoparticles naturally fulfil Kerker's condition for zero scattering into the backward direction of the incident plane waves. Secondly, a dual dipolar nanoparticle placed in a general electromagnetic field will induce an electric dipole $\mathbf{p}=\alpha_{\mathrm{e}}\mathbf{E}_{\mathrm{in}}$ and a magnetic dipole $\mathbf{m}=\alpha_{\mathrm{m}}\mathbf{H}_{\mathrm{in}}$ that preserves the local helicity of the incident field $\mathbf{E}_{\mathrm{in}}$ and $\mathbf{H}_{\mathrm{in}}$ at the origin of the particle. In the case of illumination by a circularly polarised plane wave, the helicity of the incident beam $\mathscr{H}_{\mathrm{in}}=\pm1$ is preserved. The induced electric and magnetic dipoles inside such a dual dipolar nanoparticle form a helical source that scatters light throughout the entire angular spectrum with the same helicity as the incident helical beam: $\mathscr{H}(k_x,k_y)=\mathscr{H}_{\mathrm{in}}=\pm1$.

Illustrated in Fig. \ref{fig3:HS1}(a), beam 1 along the direction $\mathbf{k}=-k_0\hat{\bm y}$ has a helicity of $\mathscr{H}_{\mathrm{in}}=-1$ which induces a helical dipolar source $\mathbf{p}_1=-i\hat{\mathbf{x}}-\hat{\mathbf{z}}$ and $\mathbf{m}_1=ic_0 \mathbf{p}_1$ that generates purely $\hat{\mathbf{e}}_{-}$ polarised scattering field over the entire angular spectrum. Beam 2 along the direction $\mathbf{k}=k_0\hat{\bm x}$ has a helicity of $\mathscr{H}_{\mathrm{in}}=+1$ which induces a helical dipolar source $\mathbf{p}_2=i\hat{\mathbf{y}}-\hat{\mathbf{z}}$ and $\mathbf{m}_2=-ic_0 \mathbf{p}_2$ that generates purely $\hat{\mathbf{e}}_{+}$ polarised scattering field over the entire angular spectrum. As a result, in the scattering angular spectrum of the nanoparticle illuminated by these two beams, the coefficient $E_{-}$ is purely determined by beam 1 and the coefficient $E_{+}$ is purely determined by beam 2. Further taking into account Kerker's condition for zero backscattering, we know that $E_{-}=0$ at $\mathbf{k}=k_0\hat{\mathbf{y}}$, the backward direction to beam 1, and $E_{+}=0$ along $\mathbf{k}=-k_0\hat{\mathbf{x}}$, the backward direction to beam 2. Since beam 1 and beam 2 independently determine $E_{-}=0$ and $E_{+}=0$, any linear combination $u_1\mathbf{p}_1+u_2\mathbf{p}_2$ and $u_1\mathbf{m}_1+u_2\mathbf{m}_2$ (where $u_1$ and $u_2$ can be any complex value) will meet the aforementioned requirements and will have zero scattering at the designated wavevector. One can see from Fig. \ref{fig3:HS1} that the scattering of the induced dipole $\mathbf{p}_1+\mathbf{p}_2$ and $\mathbf{m}_1+\mathbf{m}_2$ into wavevector $k_{\rho}=\sqrt{2}k_0$ at $\varphi=135^{\circ}$ in the upper half space is zero, as desired. Meanwhile, in the lower half space, there is no helicity singularity and no zero scattering wavevectors exist. 

\begin{figure}[!ht]
\centering
\includegraphics[width=0.45\textwidth]{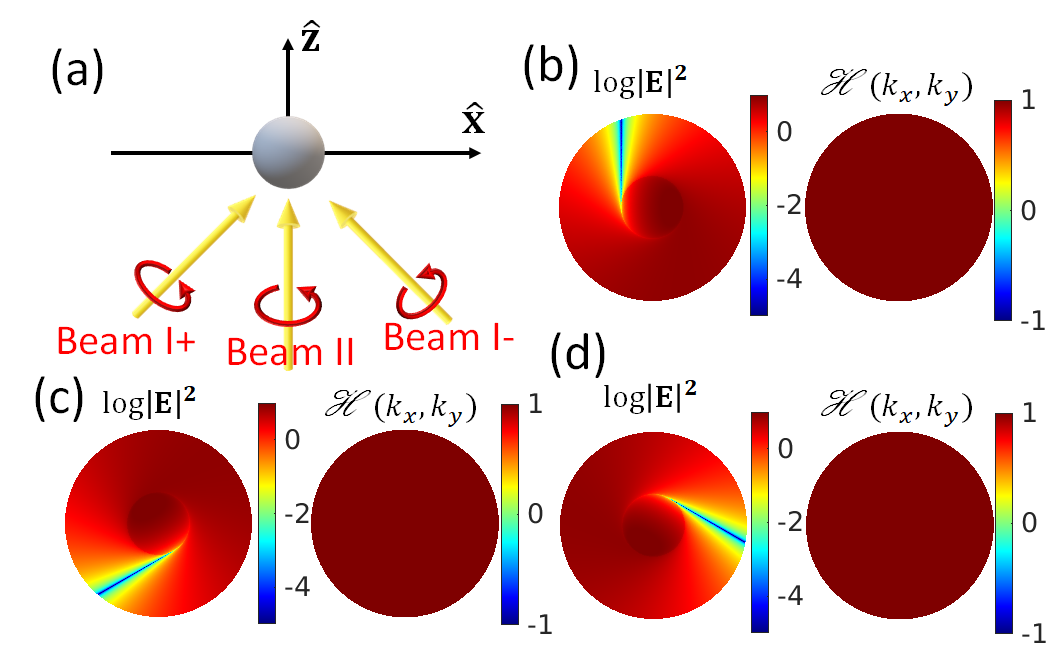}
\caption{(a). A dual dipolar nanoparticle illuminated by three helical beams. The oblique beam $\mathrm{I}+$ and beam $\mathrm{I}-$ combined induce a collinear helical dipole $\mathbf{p}_{\mathrm{I}}=\mathrm{e}^{i\Delta\phi}\hat{\mathbf{z}}$ and $\mathbf{m}_{\mathrm{I}}=-ic_0\mathbf{p}_{\mathrm{I}}$ while beam II induces a helical dipole $\mathbf{p}_{\mathrm{II}}=\hat{\mathbf{x}}+i\hat{\mathbf{y}}$ and $\mathbf{m}_{\mathrm{II}}=-ic_0 \mathbf{p}_{\mathrm{II}}$; The scattering energy and helicity angular spectrum in the upper half space for the total induced dipole $\mathbf{p}_{\mathrm{I}}+\mathbf{p}_{\mathrm{II}}$ and $\mathbf{m}_{\mathrm{I}}+\mathbf{m}_{\mathrm{II}}$ for (b) $\Delta\phi=0$ rad, (c) $\Delta\phi=2\pi/3$ rad and (d) $\Delta\phi=4\pi/3$ rad.}
\label{fig4:HS2}
\end{figure}
Angular tuning of directional scattering has been proposed \cite{LWeiOL2017} with a pair of spinning magnetic dipole $\mathbf{m}=c_0(\hat{\mathbf{x}}+i\hat{\mathbf{y}})$ and axial electric dipole $\mathbf{p}=\mathrm{e}^{i\phi_0}\hat{\mathbf{z}}$. A helical version of such dipolar source with $\mathbf{p}=\hat{\mathbf{x}}+i\hat{\mathbf{y}}-\mathrm{e}^{i\Delta\phi}\hat{\mathbf{ z}}$ and $\mathbf{m}=-ic_0\mathbf{p}$ extends the zero scattering at a single $k$ point on the $k_{\rho}=k_0$ circle to all the evanescent wavevectors on the corresponding half tangent line. This helical source can be induced inside a dual dipolar nanoparticle by a three beam excitation configuration as illustrated in Fig. \ref{fig4:HS2}(a). All three beams have the same helicity $\mathscr{H}_{\mathrm{in}}=+1$. Beam II along $k_0\hat{\mathbf{z}}$ induces a helical dipole $\mathbf{p}_{\mathrm{II}}=\hat{\mathbf{x}}+i\hat{\mathbf{y}}$ and $\mathbf{m}_{\mathrm{II}}=-ic_0 \mathbf{p}_{\mathrm{II}}$. The oblique beam I$+$ and beam I$-$ have identical amplitude and phase along wavevectors $\mathbf{k}=\pm k_x\hat{\mathbf{x}}+k_z\hat{\mathbf{z}}$ ($0<k_x\le k_0$ and $k_z=\sqrt{k_0^2-k_x^2}$) and induce a collinear helical dipole $\mathbf{p}_{\mathrm{I}}=\mathrm{e}^{i\Delta\phi}\hat{\mathbf{z}}$ and $\mathbf{m}_{\mathrm{I}}=-ic_0\mathbf{p}_{\mathrm{I}}$, where the phase factor $\mathrm{e}^{i\Delta\phi}$ is introduced by a phase difference $\Delta\phi$ between beam II and beams I$\pm$. The coefficient $E_{+}$ on the $k_{\rho}=k_0$ circle of the total induced dipole can be calculated from Eq. (\ref{eq:dipoleangularspectrumhelical}): $\mathrm{E}_{+}\propto(\mathbf{p}_{\mathrm{I}}+\mathbf{p}_{\mathrm{II}})\cdot\hat{\mathbf{e}}_{-}=\mathrm{e}^{i\varphi}+\mathrm{e}^{i\Delta\phi}$, showing that scattering into the wavevector of $\varphi=\pi+\Delta\phi$ on the $k_{\rho}=k_0$ circle is zero. As the scattering field preserves the helicity of the incident beams $\mathscr{H}_{\mathrm{in}}=+1$, the evanescent fields of wavevectors on the tangent line starting from this wavevector along $-\hat{\bm\upvarphi}$ must also be zero. As shown in Fig. \ref{fig4:HS2}, by sweeping the phase difference of the excitation beams $\Delta\phi$ from 0 to $2\pi$, the helicity of the scattering wave remains constant, while the zero scattering wavevectors sweep over the $k_{\rho}=k_0$ circle and the entire evanescent angular spectrum in momentum space.

\section{Conclusion}
We have shown that the helicity-dependent half tangent line feature of helical polarizations is a general universal property of helical evanescent waves. This momentum space geometric structure reveals itself as a remarkable near-field and far-field relationship, revealing how the entire evanescent angular spectrum can be known from the electromagnetic fields on the propagating $k_{\rho}=k_0$ circle. This feature plays a key role in the helicity angular spectrum of dipolar sources that exhibit near-field directionality: it allows us to reinterpret the null-radiation $k$ point wavevectors as being helicity singularities. A simple method consisting of structured helical beams and helicity preserving dual nanoparticles is devised from the extraordinary property of helical evanescent waves and shown as a powerful tool to design, excite and gain full control of near-field directional sources via structured far field plane wave superposition. Due to the generality of this property of helical evanescent waves, the result discussed in the current work is expected to have great impact on research involving chiral light matter interaction, directional light coupling and topological photonic structures among many others.

\section*{Acknowledgement}
This work is supported by European Research Council Starting Grant No. ERC-2016-STG-714151-PSINFONI.

\appendix
\setcounter{table}{0}
\renewcommand{\thetable}{A\arabic{table}}
\section{Representation of angular spectrum of dipolar sources in TM-TE and helical bases}
As shown in Eq. (\ref{eq:psbasis}) and Eq. (\ref{eq:helicalbasis}), the field angular spectrum $\mathbf{E}(k_x,k_y)$ can be decomposed both into TM-TE basis $\left\lbrace \hat{\mathbf{e}}_{s}, \hat{\mathbf{e}}_{p} \right\rbrace$ and into helical basis $\left\lbrace \hat{\mathbf{e}}_{+}, \hat{\mathbf{e}}_{-} \right\rbrace$:
\begin{equation}\label{append:fieldbasis}
\mathbf{E}(k_x,k_y)=E_p\hat{\mathbf{e}}_p+E_s\hat{\mathbf{e}}_s=E_{+}\hat{\mathbf{e}}_{+}+E_{-}\hat{\mathbf{e}}_{-},
\end{equation}
where the basis vectors $\hat{\mathbf{e}}_{p}$, $\hat{\mathbf{e}}_{s}$ and $\hat{\mathbf{e}}_{\pm}$ are given in Eq. (\ref{pshelicalvector}) with the relationship:
\begin{equation}\label{append:pshelicalbasis}
\hat{\mathbf{e}}_{\pm}(k_x,k_y)=\frac{1}{\sqrt{2}}(\hat{\mathbf{e}}_p\pm i\hat{\mathbf{e}}_s).
\end{equation}
From Eq. (\ref{append:fieldbasis}) and Eq. (\ref{append:pshelicalbasis}), it is easy to devise the relationship between the field coefficients $E_p$, $E_s$ and $E_{\pm}$:
\begin{equation}\label{append:coeficientpshelical}
E_{\pm}=\frac{1}{\sqrt{2}}(E_p\mp i E_s).
\end{equation}
The TM-TE basis vectors $\left\lbrace \hat{\mathbf{e}}_{s}, \hat{\mathbf{e}}_{p} \right\rbrace$, as discussed in Ref. \cite{MFPicardiPRB2017}, have the following properties:
\begin{align}\label{append:psbasisprop}
&\hat{\mathbf{e}}_{p}\cdot\hat{\mathbf{e}}_{p}=\hat{\mathbf{e}}_{s}\cdot\hat{\mathbf{e}}_{s}=1,\\\nonumber
&\hat{\mathbf{k}}\cdot\hat{\mathbf{e}}_{p}=\hat{\mathbf{k}}\cdot\hat{\mathbf{e}}_{s}=0,\\\nonumber
&\hat{\mathbf{e}}_{p}\cdot\hat{\mathbf{e}}_{s}=0,\\\nonumber
&\hat{\mathbf{e}}_{p}\times\hat{\mathbf{e}}_{s}=\hat{\mathbf{k}},
\end{align} 
where $\hat{\mathbf{k}}=\mathbf{k}/k_0$ and $\hat{\mathbf{k}}\cdot\hat{\mathbf{k}}=1$. These properties are valid for both propagating ($k_{\rho}\le k_0$) and evanescent ($k_{\rho}>k_0$) wavevectors. Note that the properties in Eqns (\ref{append:psbasisprop}) are very similar to the usual definition of orthonormal basis, but importantly, the orthogonality condition is missing a complex conjugation. This means that we can treat the vectors $\{\hat{\mathbf{e}}_{p},\hat{\mathbf{e}}_{s}\}$ as a special type of orthonormal basis, different to the traditional definition, as long as we remember that the vector components will be given via dot product without complex conjugate: 
\begin{equation}\label{append:innerproductps}
E_p=\mathbf{E}\cdot\hat{\mathbf{e}}_p,\,\,\,E_s=\mathbf{E}\cdot\hat{\mathbf{e}}_s,
\end{equation}

We can find similar properties for the helical basis. Based on Eqns (\ref{append:psbasisprop}), the following properties of the helical basis vectors defined in Eq. (\ref{append:pshelicalbasis}) can be devised:
\begin{align}\label{append:helicalbasisprop}
&\hat{\mathbf{e}}_{+}\cdot\hat{\mathbf{e}}_{+}=\hat{\mathbf{e}}_{-}\cdot\hat{\mathbf{e}}_{-}=\frac{1}{2}\left(\hat{\mathbf{e}}_{p}\cdot\hat{\mathbf{e}}_{p}-\hat{\mathbf{e}}_{s}\cdot\hat{\mathbf{e}}_{s}\right)=0,\\\nonumber
&\hat{\mathbf{k}}\cdot\hat{\mathbf{e}}_{\pm}=\frac{1}{\sqrt{2}}\left(\hat{\mathbf{k}}\cdot\hat{\mathbf{e}}_{p}\pm i\hat{\mathbf{k}}\cdot\hat{\mathbf{e}}_{s}\right)=0,\\\nonumber
&\hat{\mathbf{e}}_{+}\cdot\hat{\mathbf{e}}_{-}=\frac{1}{2}\left(\hat{\mathbf{e}}_{p}\cdot\hat{\mathbf{e}}_{p}+\hat{\mathbf{e}}_{s}\cdot\hat{\mathbf{e}}_{s}\right)=1,\\\nonumber
&\hat{\mathbf{e}}_{+}\times\hat{\mathbf{e}}_{-}=-i\hat{\mathbf{k}}. 
\end{align} 
Just like TM-TE basis, these properties of helical basis are valid for both propagating and evanescent wavevectors.\\
These properties Eqns. (\ref{append:helicalbasisprop}) are very different from the usual properties of an orthonormal basis. However, they still allow us to find the components of a vector in helical basis via simple dot products with the basis vectors $\left\lbrace \hat{\mathbf{e}}_{+}, \hat{\mathbf{e}}_{-} \right\rbrace$, as long as we remember to swap the basis vector to the opposite helicity component that we wish to find:
\begin{equation}\label{append:innerproducthelcial}
E_{+}=\mathbf{E}\cdot\hat{\mathbf{e}}_{-},\,\,\,E_{-}=\mathbf{E}\cdot\hat{\mathbf{e}}_{+}.
\end{equation}
Therefore, we can also consider this basis as being orthonormal in a non-traditional sense. Next we apply these results to the angular spectrum of a dipolar source. \\

The TM and TE field coefficients $E_p$ and $E_s$ of the angular spectrum $\mathbf{E}(k_x,k_y)$ of the radiation field by a source combination of an electric dipole $\mathbf{p}$ and a magnetic dipole $\mathbf{m}$ are expressed in Eqns (\ref{eq:dipoleangularspectrumsp}) following the derivation detailed in Ref. \cite{MFPicardiPRB2017}.\\
The field coefficients $E_{\pm}$, as a result of projection of the radiation field to the helical basis $\left\lbrace \hat{\mathbf{e}}_{+}, \hat{\mathbf{e}}_{-} \right\rbrace$, can be derived based on the properties in Eqn. (\ref{append:helicalbasisprop}) as follows:
\begin{align}\label{append:fieldhelicalbasis}
&E_{+}\propto\left(\mathbf{p}-\hat{\mathbf{k}}\times\frac{\mathbf{m}}{c_0}\right)\cdot\hat{\mathbf{e}}_{-}=\left(\mathbf{p}+ i\frac{\mathbf{m}}{c_0}\right)\cdot\hat{\mathbf{e}}_{-},\\\nonumber
&E_{-}\propto\left(\mathbf{p}-\hat{\mathbf{k}}\times\frac{\mathbf{m}}{c_0}\right)\cdot\hat{\mathbf{e}}_{+}=\left(\mathbf{p}- i\frac{\mathbf{m}}{c_0}\right)\cdot\hat{\mathbf{e}}_{+}.
\end{align} 
Alternatively, the same expression of field coefficients $E_{\pm}$ in helical basis as shown in Eqns. (\ref{append:fieldhelicalbasis}) and Eq. (\ref{eq:dipoleangularspectrumhelical}) can be obtained following Eq. (\ref{append:coeficientpshelical}) and the field angular spectrum in TM-TE basis in Eqns. (\ref{eq:dipoleangularspectrumsp}). 
\section{Physical properties of a propagating wave}
In comparison with evanescent waves as shown in Table \ref{table1}, the same set of physical properties for a propagating wave along the direction $\hat{\mathbf{k}}=\mathbf{k}/k_0$ under different polarization bases are listed in TABLE \ref{table2}.
\begin{table}[!htbp]
\centering
\caption{Helicity $\mathscr{H}$, normalised Poynting vector $\hat{\mathbf{P}}$ and spin vectors $\hat{\mathbf{s}}_E$ of the electric field and $\hat{\mathbf{s}}_H$ magnetic field of a propagating plane wave for $\hat{\mathbf{e}}_p$, $\hat{\mathbf{e}}_s$ and $\hat{\mathbf{e}}_{\pm}$ polarised fields.}
\begin{tabular}{|c|c|c|c|}
\hline
&$\hat{\mathbf{e}}_p$ (TM)&$\hat{\mathbf{e}}_s$ (TE) &$\hat{\mathbf{e}}_{\pm}$ (Helical) \\
\hline
$\hat{\mathbf{s}}_E$&0&0&$\pm\hat{\mathbf{k}}$\\
\hline
$\hat{\mathbf{s}}_H$&0&0&$\pm\hat{\mathbf{k}}$\\
\hline
$\hat{\mathbf{P}}$&$\hat{\mathbf{k}}$&$\hat{\mathbf{k}}$&$\hat{\mathbf{k}}$\\
\hline
$\mathscr{H}$&0&0&$\pm1$\\
\hline
\end{tabular}
\label{table2}
\end{table}

%
\end{document}